# Edge Exposure as the Trigger for Structural Instability in LP-N and HLP-N

Guo Chen[1,2], Xianlong Wang[1,2*]

[1]*Key Laboratory of Materials Physics, Institute of Solid State Physics, HFIPS, Chinese Academy of Sciences, Hefei 230031, China*

[2]*University of Science and Technology of China, Hefei 230026, China*

___________

[*]Author to whom all correspondence should be addressed: xlwang@theory.issp.ac.cn,

# ABSTRACT

LP-N and HLP-N are promising high-energy-density materials. However, these materials synthesized under high pressure cannot be maintained stable at ambient conditions. The mechanism behind their instability remains unclear. Our research, based on first-principles calculations and ab initio molecular dynamics simulations, reveals that while not edge exposed, LP-N and HLP-N exhibit substantial structural, dynamic, and mechanical stability under ambient conditions. The stability of HLP-N is governed by an interlocking mechanism, which becomes ineffective upon exposure of the edges, leading to internal breakdown. As a result, H saturated adsorption has no impact on it. In contrast, LP-N benefits modestly from H saturated adsorption due to its edge-initiated dissociation. The interlocking mechanism offer valuable insights into the design of new materials.

**INTRODUCTION**

Energetic materials release energy through bond breaking and reformation in chemical reactions, with significant potential in applications such as energy storage, propellants, and explosives [1,2]. Traditional explosives rely on the breaking of chemical bonds like C-N and C-O [3]. In contrast, polymeric nitrogen (PN) can achieve complete reaction without an oxidizer, forming a high-density network of single bonds (N-N, bonding energy of 160 kJ/mol) and double bonds (N=N, bonding energy of 418 kJ/mol). Upon decomposition into non-polluting $N_2$ (N≡N, bonding energy of 954 kJ/mol) [4], the substantial difference in bond energies results in a large energy release. Consequently, PN is considered a high-energy-density material (HEDM).

Nitrogen gas is composed of strongly triple bonds, which are extremely difficult to dissociate under ambient pressure. Under pressure, the structural diversity of nitrogen is significantly enhanced as a result of compressed atomic spacing, electronic structure reconstruction, and coordination pattern modification [5]. However, the complexity of nitrogen's phase diagram is also attributed to its five valence electrons in the outermost shell. For example, at low pressure, the intermolecular dissociation transitions in nitrogen involve the rearrangement of triple bonds, leading to various molecular solids, such as α-N, β-N, γ-N, δloc-N, δ-N, ε-N, λ-N, ι-N, θ-N, ζ-N, κ-N, η-N [6–13]. At higher pressure, the compression of molecular spacing induces electron delocalization, thereby lowering the energy barrier of the covalent triple bonds. Notably, In 1985, McMahan *et. al.* first theoretically predicted the existence of PN composed solely of N-N covalent bonds at high pressure [14].

Despite theoretical predictions of numerous PN phases with diverse dimensional structures [e.g., zero-dimensional (0D) cage-like [15], one-dimensional (1D) chain-like [16], two-dimensional (2D) layered [17–20], and three-dimensional (3D) network structures [21]], only a limited number of these phases have been successfully synthesized under ambient or extreme conditions to date. For instance, the 3D network-structured cubic gauche nitrogen (cg-N) synthesized at 110 GPa, while 2D layered structures such as layered polymeric nitrogen (LP-N), hexagonal layered polymeric nitrogen (HLP-N), black phosphorus nitrogen (BP-N), and post-layered-polymeric nitrogen (PLP-N) synthesized at 126 GPa, 244 GPa, 146 GPa, and 161 GPa, respectively. Notably, these high-pressure-synthesized PNs cannot stabilize under ambient pressure, with their structural integrity maintained only via quenching processes to 40–60 GPa [17–21]. This has made achieving stable synthesis of PN under ambient conditions a central research challenge. In response, researchers have developed innovative strategies including surface modification, nano-confinement, and doping methods. For example, at ambient pressure, carbon doping and hydrogen saturation of surface dangling bonds were demonstrated to enhance the kinetic stability of cg-N [22]; Furthermore, phosphorus doping was proven to enchance the stability of BP-N to 0 GPa [23]. Charge transfer and vdW confinement effect were proposed to stabilize layered LP-N and HLP-N under ambient conditions [24]. Nevertheless, the instability mechanisms of LP-N and HLP-N are still not clear.

Our research explored the structural, dynamics, mechanical stability of LP-N and HLP-N under low pressure using first-principles calculations and AIMD simulations. We discovered edge exposed significantly reduces their stability. In contrast to LP-N, which starts breaking down from its edges, HLP-N is susceptible to internal collapse due to the loss of its interlocking mechanism. Furthermore, H saturated apsorption can improve the stability of LP-N, it has no such effect on HLP-N.

**METHODS**

The computational methodology for investigating phonons, elastic constants, and detonation performance involves employing density functional theory (DFT) within the framework of the all-electron projector augmented wave (PAW) method, as realized in the VASP software package [25,26]. The Perdew-Burke-Ernzerhof (PBE) [27] exchange-correlation functional under the generalized gradient approximation (GGA) [28,29] was utilized. A plane-wave kinetic energy cutoff of 520 eV was employed. For phonon dispersion and elastic constants, total enthalpies converged to 1E-8 eV, while for detonation performance, convergence was set to 1E-6 eV between consecutive steps enthalpiles change. The Monkhorst-Pack (MP) scheme with a k-grid spacing of 0.15 Å$^2$ was used to sample the reciprocal space. Semi-empirical van der Waals interactions were included via the DFT-D3 correction [30].

Phonon dispersions were calculated using the PHONOPY code based on the finite displacement method, with atoms displaced by ±0.015 Å to construct force constant matrices [31,32]. The elastic constant matrix was computed via the stress-strain approach [33], and the mechanical stability was assessed according to the Born-Huang criteria [34]. Bulk modulus $B$, shear modulus $G$, Poisson's ratio $v$, and Vickers hardness $Hv$ were derived from the elastic constant matrix using the Voigt-Reuss-Hill method [35]. For energy density calculations, BP-N and α-N$_2$ were selected as ground-state phases. Detonation velocity $V_d$ and detonation pressure $P_d$ were determined using the Kamlet-Jacobs empirical equations [36]. All formulas are provided in the supplementary material.

Structural relaxation and molecular dynamics simulations were performed using CP2K [37], employing density functional theory (DFT) within the The Gaussian and plane wave (GPW) approach [38]. Calculations used the DZVP-MOLOPT-SR-GTH basis set with a plane-wave cutoff energy of 400 Ry and a relative cutoff of 55 Ry. The convergence criterion for the density matrix was set to $1.0 \times 10^{-6}$ eV. Van der Waals corrections were applied via the DFT-D3 method [30]. A k-grid spacing of 0.30 Å$^{-1}$ was applied to all slabs. For asymmetric slabs or those with dangling bonds, dipole corrections were employed along the z-axis. For structural relaxation, constraints were placed on the maximum ($\leq 4.5 \times 10^{-4}$ eV/Å) and RMS force ($\leq 3 \times 10^{-4}$ eV/Å). Molecular dynamics simulations under NpT conditions utilized the canonical sampling through velocity rescaling (CSVR) thermostat [39], enabling relaxation of cell dimensions in the XY plane. The slab configuration details are provided in the supplementary material.

**RESULTS AND DISCUSSION**

LP-N and HLP-N adopt the Pba2 (No. 32, Z = 16) and P42bc (No. 106, Z = 32) space groups for their crystal structures. LP-N exhibits thermodynamic stability at static conditon between 188-320 GPa, while HLP-N remains metastable at all pressures. The Fig.1 displays their crystal structures at 0 GPa, with each layer being structurally equivalent. Nevertheless, HLP-N exhibits an AB layer configuration, with adjacent layers being rotated 90° horizontally relative to each other. Calculations show LP-N and HLP-N can stay structurally stable at 0 GPa with no imaginary frequencies, indicating the possibility of achieving their metastable state under ambient conditions.

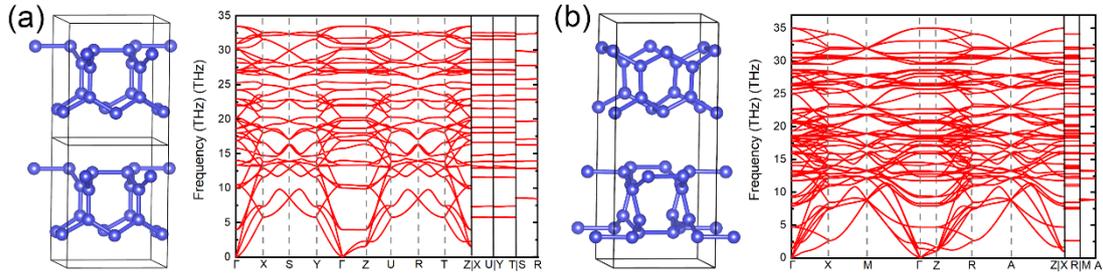

**Fig. 1.** Structures and phonon dispersions of (a) the 1x1x2 supercell of LP-N and (b) HLP-N at 0 GPa.

In addition to phonon dispersion, the mechanical properties of LP-N and HLP-N at 0 GPa were analyzed. As shown in Table 1, their elastic constants are highly similar and meet the Born criteria [34], confirming the mechanical stability of LP-N and HLP-N at 0 GPa. The relationship $C_{11} = C_{22} > C_{33}$ indicates that both materials are isotropic along the x and y axes with greater stiffness than along the z axis, making them more compressible within layers. Their bulk moduli $G$ exceed the shear moduli $B$, signifying higher resistance to volumetric deformation. The bulk-to-shear modulus ratios $B/G$ (1.113 for LP-N and 1.080 for HLP-N) are below 1.75, and the Poisson's ratios $v$ (0.154 for LP-N and 0.146 for HLP-N) are below 0.26, indicating brittle materials with some plastic behavior. Finally, their Vickers hardness $H_v$ values exceed 20 GPa (22.092 GPa for LP-N and 22.533 GPa for HLP-N), classifying them as hard materials.

**Table 1.** Elastic constants (GPa) of LP-N and HLP-N at 0 GPa.

| Phase | $C_{11}$ | $C_{22}$ | $C_{33}$ | $C_{44}$ | $C_{55}$ | $C_{66}$ | $C_{12}$ | $C_{13}$ | $C_{23}$ |
|---|---|---|---|---|---|---|---|---|---|
| LP-N | 721.244 | 721.244 | 39.193 | 18.186 | 18.186 | 268.961 | 35.414 | 5.065 | 5.065 |
| HLP-N | 678.765 | 678.767 | 39.618 | 15.671 | 15.666 | 285.487 | 21.143 | 4.967 | 4.965 |

**Table 2.** The calculated bulk modulus *B*, shear modulus *G*, bulk-to-shear modulus ratios *B/G*, Poisson's ratio *v* and Vickers hardness $H_v$ of LP-N and HLP-N at 0 GPa.

| Phase | $B_{VRH}$(GPa) | $G_{VRH}$(GPa) | B/G | v | $H_v$(GPa) |
|---|---|---|---|---|---|
| LP-N | 105.543 | 94.849 | 1.113 | 0.154 | 22.092 |
| HLP-N | 99.300 | 91.982 | 1.080 | 0.146 | 22.533 |

Due to composition entirely of N-N single bonds, LP-N and HLP-N potentially possess ultra-high energy density. We calculated their detonation performance at 0 GPa and compared them with the calculated results of cg-N, as well as the experimental data of 2,4,6-trinitrotoluene (TNT) and 1,3,5,7-tetranitro-1,3,5,7-tetrazocane (HMX) [40], as shown in the Fig. 2. To address the solid-to-gas transition, we introduced a chemical potential-based correction. The energy drop from solid to gaseous $N_2$, estimated by adding the 298 K chemical potential to the 0 K enthalpy, is 0.50 eV/$N_2$ [41]. As shown in Fig. 3, the performance of LP-N, HLP-N, and cg-N significantly exceeds that of traditional energetic materials like TNT and HMX. Specifically, the energy densities of LP-N and HLP-N are 12.04 kJ/g and 12.67 kJ/g, respectively, which are higher than cg-N (10.8 kJ/g) and far exceed those of TNT (4.3 kJ/g) and HMX (5.7 kJ/g). The values for detonation velocity ($V_d$) and pressure ($P_d$) show the following order: LP-N > cg-N > HLP-N. Specifically, the detonation velocities of LP-N and HLP-N are 26.75 km/s and 23.24 km/s, respectively, which are 3.84 times and 3.33 times that of TNT (6.97 km/s).

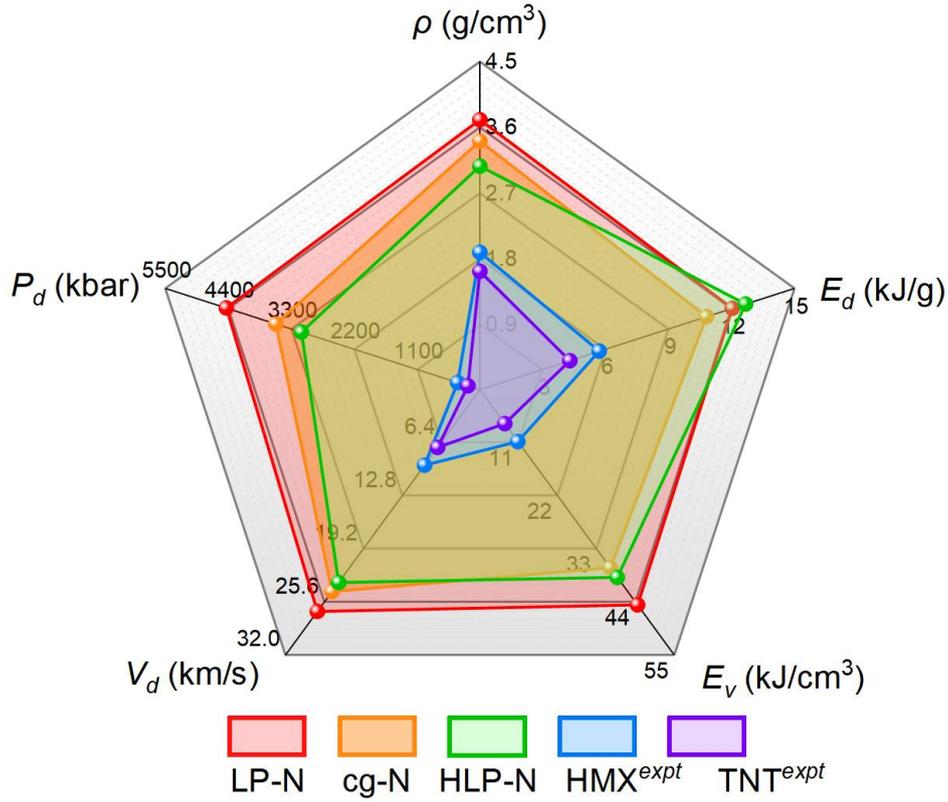

Fig. 2. The calculated properties for LP-N and HLP-N include density $\rho$, gravimetric energy density $E_d$, volumetric energy density $E_v$, detonation velocity $V_d$, and detonation pressure $P_d$. For comparison, the calculated results for cg-N at the same levels with LP-N and HLP-N, along with experimental data for TNT and HMX explosives, are also provided. The superscript '*expt*' denotes experimental data.

Despite computational indications of LP-N and HLP-N's stability (structural, dynamic, and mechanical) at 0 GPa and the high denotation perfermance, experimental high-pressure synthesized structures cannot be retained at ambient pressure, and the underlying mechanism of this instability has yet to be elucidated. Previous calculations and experiments show that cg-N's instability is due to surface instability [22,42]. This suggests that the instabilities observed in LP-N and HLP-N may be caused in their edge structures, since in reality, they are not perfect crystals but rather expose edges. On this basis, we first analyzed LP-N's edges, with the stability of its slabs including (001)-surface and different sub-edges [Miller indices of (100), (110), and (111), with or without H saturated adsorption] at 0 GPa and 0 K, as shown in Fig. 3.

Most pristine slabs, such as the (100b), (100c), (110a), and (110b), are unstable at 0 GPa. These slabs typically decompose to release $N_2$, with the (100b)-slab additionally forming $N_5$ rings. However, H saturated adsorption can stabilize all edges of these slabs at 0 GPa and static condition. The initial edges of H saturated adsorption are shown in Supplementary Materials Fig. S2. Although for those stable pristine slabs with edge exposed, all undergo reconstruction. For example, the layer of the (100a)-slab bends, whereas the (110b) and (111b) slabs achieve stability through the process of partial $N_2$ dissociation followed by reconstruction into N-chains. Since some cg-N surfaces reconstruct to form $N_3^-$ and can be more readily synthesized at ambient pressure using azides as precursors [43], LP-N edges, which reconstruct into $N_3^-$ and decompose into $N_5$ rings, might be more easily synthesized from azides or $N_5$ rings.

The Kleinman parameter ($\zeta$) determines the stability of structure under tensile and bending types of strain [44]. For LP-N, $\zeta=0.20$ suggests that its instability is likely driven by bond bending. e.g. (100a)-slab. Additionally, the crystal contains three distinct types of bonds, as illustrated in the Supplementary Materials Fig. S3 (a), where the bond length between N1-N2 is 1.624 Å, which significantly exceeds the other two types of bonds whose lengths are approximately 1.4 Å. Upon exposure of the edges in the (100a)-slab, bending occurs, leading to the breaking of the weaker, longer bonds. The bending in the (100a)-slab occurs under low pressure, as demonstrated in Supplementary Fig. S4, where we simulated the decompression process. It was observed that bending began as the pressure was gradually reduced. Nonetheless, H-saturated adsorption can prevent this bending and inhibit most reconstruction phenomena.

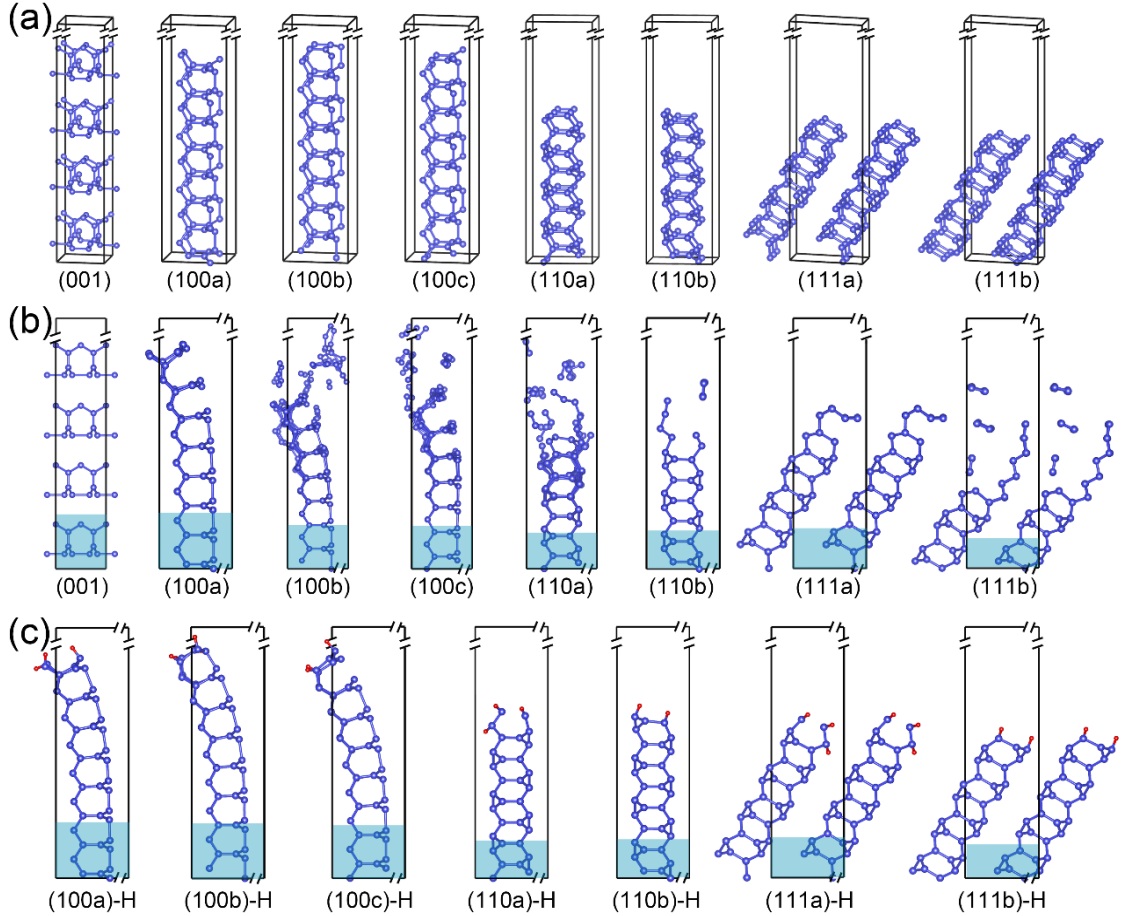

Fig. 3. Stabilities of the different LP-N slabs at 0 GPa and 0 K. (a) Initial and (b) Relaxed slabs of (100)-surface and different edges. (c) Relaxed slabs with H saturated adsorption. The breakpoints on the horizontal (vertical) axis signify the incomplete periodic structures (vacuum). Blue and red balls correspond to nitrogen and hydrogen atoms respectively.

In order to explore the thermal stability of different LP-N slabs at 0 GPa and 300K, AIMD simulations were executed, with the MSD trends as shown in Fig. 4(c). It indicates a significant upward trend for all exposed-edge slabs except for the non-exposed edge (001)-slab, suggesting the instability of LP-N at low pressure is primarily due to the edges exposed. Nevertheless, H saturated adsorption delays the onset of decomposition, exemplified by (100a)-H, (110b)-H, and (111a)-H slabs, whose MSD curves show a delayed dramatic increase compared to the unexposed countparts. This delay is attributed to the inhibition of edge instability by hydrogen saturated adsorption.

However, ultimate decomposition occurs from the sides due to edge exposed, as shown in Figs. 4(a)-(b). Moreover, LP-N with H saturated adsorption can even decompose internally, particularly for (110) slabs, with the (110b)-H shown in Fig. 4(c).

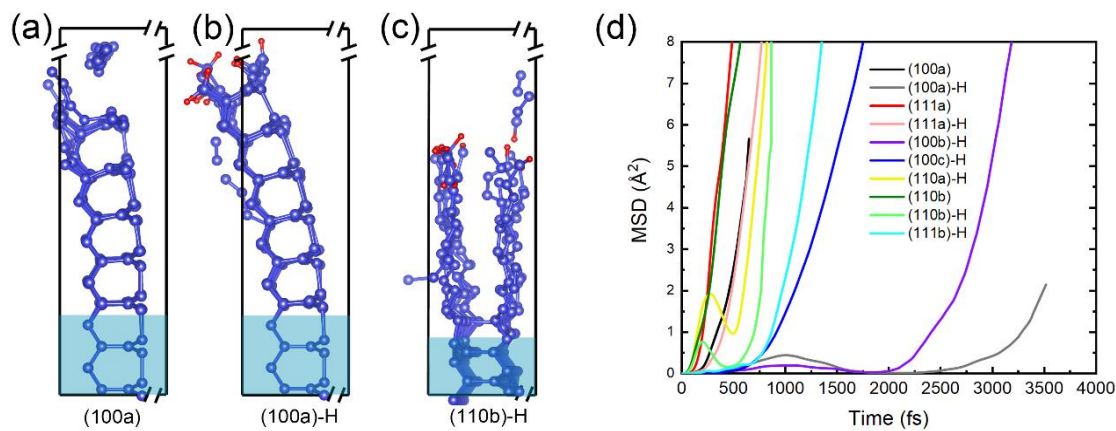

Fig. 4. Dynamic behavior analysis of various slabs at 0 GPa and 300 K. (a)-(d) The structures of partial decomposition of the (100a), (100a)-H, and (110b)-H slabs. (d) The MSD changes over time. The breakpoints on the horizontal (vertical) axis signify the incomplete periodic structures (vacuum). Blue and red balls correspond to nitrogen and hydrogen atoms respectively.

To investigate the instability mechanism of HLP-N, we constructed slab models of monolayer and bilayer HLP-N slabs without edge exposed and performed relaxation at 0 GPa and 0 K. The results, shown in Fig. 5, indicate that while the monolayer structure becomes unstable along the x-axis, the bilayer structure remains stable. The instability in the monolayer is attributed to the breaking of its longest bond (1.640 Å), as shown in supplementary material Fig. 3(b), which significantly exceeds the other two types of bonds ($\approx$1.4 Å).

The relaxed bilayer structure has identical lengths along both the x and y axes (4.64 Å), whereas the monolayer expands to 7.60 Å along the x-axis and contracts to 4.47 Å along the y-axis. The contraction in the perpendicular direction to decomposition in monolayers could be responsible for the stability seen in bilayer systems. Concerning the bilayer HLP-N slab, as exemplified in Fig. 5(e), the lower layer exhibits a tendency to decompose along the x-axis but concurrently generates a contracting force along the y-axis, which inhibits the upper layer's decomposition prone along this axis. Similarly, the upper layer's contraction along the x-axis could inhibit the lower layer's decomposition. We term this behavior an interlocking mechanism, a principle that can be harnessed for stabilizing existing structures or designing novel ones.

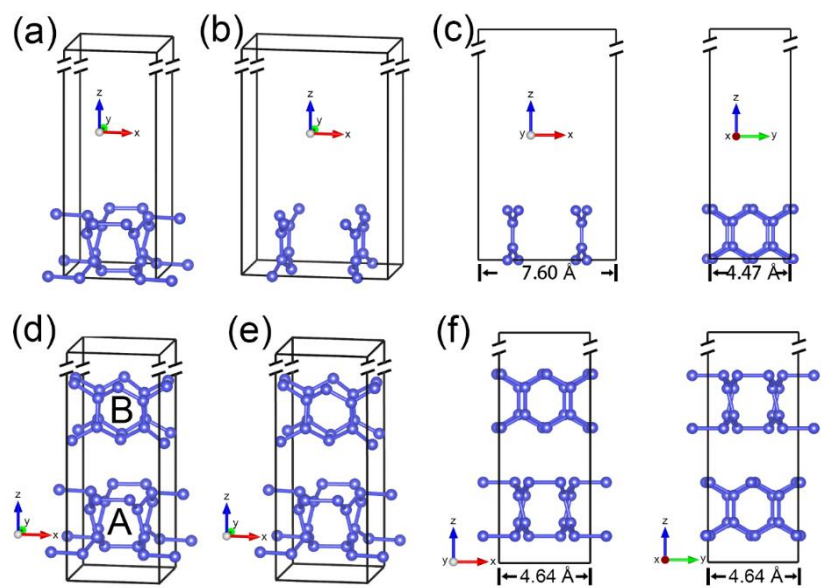

Fig. 5. Stability of monolayer and bilayer HLP-N slabs at 0 GPa and 0 K. (a) Initial and (b) relaxed monolayer configuration. (c) Multiple views of the relaxed monolayer configuration. (d) Initial, and (e) relaxed bilayer configuration. (f) Multiple views bilayer configuration. The breakpoints on the vertical axis signify the incomplete vacuum.

For validating the interlocking mechanism, the stability of HLP-N (100a)-slab (with and without H saturated adsorption) are investigated at 0 GPa and static condition, as shown in the Figs. 6(a)-(c). The results indicate their instability. This is because with the edge exposed, layer A, which is exposed to the vacuum, cannot benefit from layer B. Furthermore, H saturated adsorption does not improve the stability of HLP-N, as it decomposes not starting from the edges but due to destabilization of the longest bonds, leading to overall structural collapse.

AIMD simulations were performed at 0 GPa and 300 K, and the curves of MSD

with different directions over time are shown in Fig. 6(d). The MSD reveals in the initial stage, the atomic displacement along the z-axis is greater than x-axis due to the vacuum layer along the z-axis causing Layer A to decompose first, thereby generating stress along the x-axis which temporarily stabilizes Layer B. As shown in Fig. 6(e), by the 85th step of AIMD, Layer A has already decomposed along the z-axis, while Layer B remains stable. As Layer A continues to decompose, its stabilizing effect on Layer B decreases, resulting in a substantial growth of x-axis of MSD. By the 350th step of AIMD, Layer A has broken down into $N_2$ and chains along the z-axis, and Layer B starts to disintegrate along the x-axis. Hence, due to the interlocking mechanism, HLP-N exhibits stability at 0 GPa; However, this stability is compromised upon exposure of the edges, leading to a sequential decomposition starting from one layer followed by another.

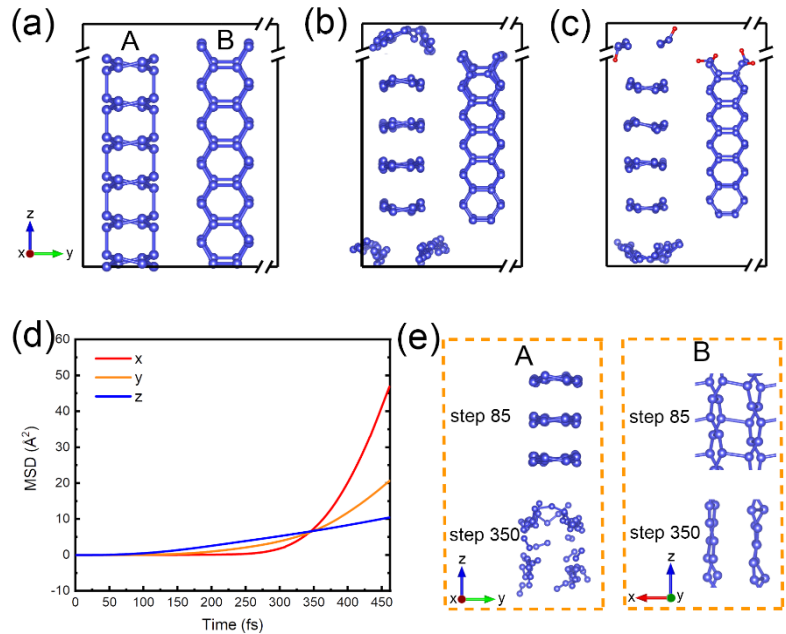

Fig. 6. Stability of edge exposed HLP-N (100a)-slab with and without H saturated adsorption. (a) Initial (100a)-slab, (b) Stability without and (c) with H saturated adsorption at 0 GPa and 0 K. (d) MSD of the (100a)-slab in various directions at 0 GPa and 300 K. (e) Structures of Layer A and Layer B of the (100a)-slab at different simulation steps at 0 GPa and 300 K. The breakpoints on the horizontal (vertical) axis signify the incomplete periodic structures (vacuum). Blue and red balls correspond to nitrogen and hydrogen atoms respectively.

**CONCLUSION**

Based on the first-principles calculations and AIMD simulations, we have investigated the stability of LP-N and HLP-N at 0 GPa. The study identified the underlying reasons for the instability of both LP-N and HLP-N, and provided strategies for enhancing the stability of LP-N.

Our results show that although LP-N and HLP-N exhibit excellent structural, dynamical, and mechanical stability at 0 GPa, once edges are exposed, their stabilities significantly deteriorate. Specifically, LP-N starts breaking down from the edges, whereas HLP-N is more prone to internal collapse. The application of H saturated adsorption offers a moderate enhancement in the stability of LP-N, yet it remains insufficient for ambient conditions. However, this method does not improve the stability of HLP-N. This is because HLP-N's stability is based on an interlocking mechanism, which breaks down when edges are exposed at 0 GPa. Utilizing this mechanism may offer possibilities for the creation of new materials.

**ACKNOWLEDGEMENTS**

This work is supported by the National Natural Science Foundation of China (NSFC) under Grant of U2030114, and CASHIPS Director's Fund (Grant No. YZJJ202207-CX). The calculations were partly performed at the Center for Computational Science of CASHIPS, the ScGrid of the Supercomputing Center and Computer Network Information Center of the Chinese Academy of Sciences, and the Hefei Advanced Computing Center.

REFERENCES

[1] C. Mailhiot, L. H. Yang, and A. K. McMahan, Polymeric nitrogen, Phys. Rev. B **46**, 14419 (1992).

[2] V. E. Zarko, Searching for ways to create energetic materials based on polynitrogen compounds (review), Combust. Explos. Shock Waves **46**, 121 (2010).

[3] Y. Xiao, L. Chen, K. Yang, J. Lu, and J. Wu, Reaction mechanism and electronic properties of host–guest energetic material CL-20/HA under high pressure by quantum-based molecular dynamics simulations, Phys. Chem. Chem. Phys. **25**, 15846 (2023).

[4] K. O. Christe, W. W. Wilson, J. A. Sheehy, and J. A. Boatz, $N_5^+$: A Novel Homoleptic Polynitrogen Ion as a High Energy Density Material, J. Am. Chem. Soc. **38**, 2004 (1999).

[5] L. N. Yakub, Polymerization in highly compressed nitrogen (Review Article), Low Temp. Phys. **42**, 1 (2016).

[6] J. Donohue, A refinement of the positional parameter in α-nitrogen, Acta Cryst. **14**, 1000 (1961).

[7] A. F. Goncharov, I. G. Batyrev, E. Bykova, L. Brüning, H. Chen, M. F. Mahmood, A. Steele, N. Giordano, T. Fedotenko, and M. Bykov, Structural diversity of molecular nitrogen on approach to polymeric states, Phys. Rev. B **109**, 064109 (2024).

[8] A. F. Goncharov, E. Gregoryanz, H. Mao, Z. Liu, and R. J. Hemley, Optical Evidence for a Nonmolecular Phase of Nitrogen above 150 GPa, Phys. Rev. Lett. **85**, 1262 (2000).

[9] E. Gregoryanz, A. F. Goncharov, R. J. Hemley, and H. Mao, High-pressure amorphous nitrogen, Phys. Rev. B **64**, 052103 (2001).

[10] M. Frost, R. T. Howie, P. Dalladay-Simpson, A. F. Goncharov, and E. Gregoryanz, Novel high-pressure nitrogen phase formed by compression at low temperature, Phys. Rev. B **93**, 024113 (2016).

[11] D. Plašienka and R. Martoňák, Transformation pathways in high-pressure solid nitrogen: From molecular N2 to polymeric cg-N, J. Chem. Phys. **142**, 094505 (2015).

[12] A. F. Schuch and R. L. Mills, Crystal Structures of the Three Modifications of Nitrogen 14 and Nitrogen 15 at High Pressure, J. Chem. Phys. **52**, 6000 (1970).

[13] R. L. Mills, B. Olinger, and D. T. Cromer, Structures and phase diagrams of N2 and CO to 13 GPa by x-ray diffraction, J. Chem. Phys. **84**, 2837 (1986).

[14] A. K. McMahan and R. LeSar, Pressure Dissociation of Solid Nitrogen under 1 Mbar, Phys. Rev. Lett. **54**, 1929 (1985).

[15] X. Wang, Y. Wang, M. Miao, X. Zhong, J. Lv, T. Cui, J. Li, L. Chen, C. J. Pickard, and Y. Ma, Cagelike Diamondoid Nitrogen at High Pressures, Phys. Rev. Lett. **109**, 175502 (2012).

[16] M. M. G. Alemany and J. L. Martins, Density-functional study of nonmolecular phases of nitrogen: Metastable phase at low pressure, Phys. Rev. B **68**, 024110 (2003).

[17] Dane Tomasino, Minseob Kim, J. Smith, and C.-S. Yoo, Pressure-Induced Symmetry-Lowering Transition in Dense Nitrogen to Layered Polymeric Nitrogen (LP-N) with Colossal Raman Intensity, Phys. Rev. Lett. **113**, 205502 (2014).

[18] D. Laniel, G. Geneste, G. Weck, M. Mezouar, and P. Loubeyre, Hexagonal Layered Polymeric Nitrogen Phase Synthesized near 250 GPa, Phys. Rev. Lett. **122**, 066001 (2019).

[19] D. Laniel, B. Winkler, T. Fedotenko, A. Pakhomova, S. Chariton, V. Milman, V. Prakapenka, L. Dubrovinsky, and N. Dubrovinskaia, High-Pressure Polymeric Nitrogen Allotrope with the Black Phosphorus Structure, Phys. Rev. Lett. **124**, 216001 (2020).

[20] C. Ji et al., Nitrogen in black phosphorus structure, Sci. Adv. **6**, eaba9206 (2020).

[21] M. I. Eremets, A. G. Gavriliuk, I. A. Trojan, D. A. Dzivenko, and R. Boehler, Single-bonded cubic form of nitrogen, Nat. Mater. **3**, 558 (2004).

[22] G. Chen, C. Niu, W. Xia, J. Zhang, Z. Zeng, and X. Wang, Route to Stabilize Cubic Gauche Polynitrogen to Ambient Conditions via Surface Saturation by Hydrogen, Chinese Phys. Lett. **40**, 086102 (2023).

[23] G. Chen, C. Zhang, Y. Zhu, B. Cao, J. Zhang, and X. Wang, Realized stable BP-N at ambient pressure by phosphorus doping, Matter Radiat. at Extremes **10**, 015801

(2025).

[24] D.-X. Wang, J. Fu, Y. Li, Z. Yao, S. Liu, and B.-B. Liu, Interception of Layered LP-N and HLP-N at Ambient Conditions by Confined Template, Chinese Phys. Lett. **41**, 036101 (2024).

[25] G. Kresse and J. Furthmüller, Efficient iterative schemes for *ab initio* total-energy calculations using a plane-wave basis set, Phys. Rev. B **54**, 11169 (1996).

[26] G. Kresse and J. Furthmüller, Efficiency of ab-initio total energy calculations for metals and semiconductors using a plane-wave basis set, Computational Materials Science **6**, 15 (1996).

[27] J. P. Perdew, K. Burke, and M. Ernzerhof, Generalized Gradient Approximation Made Simple, Phys. Rev. Lett. **77**, 3865 (1996).

[28] W. Kohn and L. J. Sham, Self-Consistent Equations Including Exchange and Correlation Effects, Phys. Rev. **140**, A1133 (1965).

[29] J. P. Perdew, Y. Wang, G. Kresse, and D. Joubert, Accurate and simple analytic representation of the electron-gas correlation energy, Phys. Rev. B **45**, 13244 (1992).

[30] S. Grimme, J. Antony, S. Ehrlich, and H. Krieg, A consistent and accurate *ab initio* parametrization of density functional dispersion correction (DFT-D) for the 94 elements H-Pu, J. Chem. Phys **132**, 154104 (2010).

[31] K. Parlinski, Z. Q. Li, and Y. Kawazoe, First-Principles Determination of the Soft Mode in Cubic $ZrO_2$, Phys. Rev. Lett. **78**, 4063 (1997).

[32] A. Togo, L. Chaput, and I. Tanaka, Distributions of phonon lifetimes in Brillouin zones, Phys. Rev. B **91**, 094306 (2015).

[33] Y. Le Page and P. Saxe, Symmetry-general least-squares extraction of elastic data for strained materials from *ab initio* calculations of stress, Phys. Rev. B **65**, 104104 (2002).

[34] M. Born, K. Huang, and M. Lax, Dynamical Theory of Crystal Lattices, Am. J. Phys. **23**, 474 (1955).

[35] L. Zuo, M. Humbert, and C. Esling, Elastic properties of polycrystals in the Voigt-

Reuss-Hill approximation, J. Appl. Crystallogr. **25**, 751 (1992).

[36] M. J. Kamlet and S. J. Jacobs, Chemistry of Detonations. I. A Simple Method for Calculating Detonation Properties of C–H–N–O Explosives, J. Chem. Phys **48**, 23 (1968).

[37] T. D. Kühne et al., CP2K: An electronic structure and molecular dynamics software package - Quickstep: Efficient and accurate electronic structure calculations, J. Chem. Phys **152**, 194103 (2020).

[38] G. Lippert, J. Hutter, and M. Parrinello, A hybrid Gaussian and plane wave density functional scheme, Mol. Phys. **92**, 477 (1997).

[39] G. Bussi, D. Donadio, and M. Parrinello, Canonical sampling through velocity rescaling, J. Chem. Phys **126**, 014101 (2007).

[40] B. M. Dobratz, LLNL Explosives Handbook: Properties of Chemical Explosives and Explosives and Explosive Simulants, No. UCRL-52997, Lawrence Livermore National Lab. (LLNL), Livermore, CA, 1981.

[41] Jr Malcolm W Chase, *NIST-JANAF Thermochemical Tables, 4th Ed* (American Institute of Physics, 1998).

[42] Y. Xu et al., Free-standing cubic gauche nitrogen stable at 760 K under ambient pressure, Sci. Adv. **10**, eadq5299 (2024).

[43] L. Wu, Y. Xu, G. Chen, J. Ding, M. Li, Z. Zeng, and X. Wang, One-step Synthesis of Cubic Gauche Polymeric Nitrogen with High Yield Just by Heating, Chinese Phys. B **33**, 126802 (2024).

[44] G. Dearnaley, Electronic Structure and Properties of Solids: The Physics of the Chemical Bond, Phys. Bull. **31**, 359 (1980).